\documentclass[twocolumn]{emulateapj}

\usepackage{amsmath}

\shortauthors{
Wong et al.
}

\shorttitle{
Bondi Accretion Flow of NGC 3115
}
\slugcomment{Astrophysical Journal Letter}

\begin{document}
\title{
Resolving the Bondi Accretion Flow toward the Supermassive Black Hole of NGC 3115 with {\it Chandra}
}

\author{
Ka-Wah Wong\altaffilmark{1},
Jimmy A. Irwin\altaffilmark{1},
Mihoko Yukita\altaffilmark{1},
Evan T. Million\altaffilmark{1},
William G. Mathews\altaffilmark{2},
and
Joel N. Bregman\altaffilmark{3}
}

\altaffiltext{1}{Department of Physics and Astronomy, University of 
Alabama, Box 870324, Tuscaloosa, AL 35487, USA
}
\altaffiltext{2}{UCO/Lick Observatory, Department of Astronomy and 
Astrophysics, University of California, Santa Cruz, CA 95064, USA
}
\altaffiltext{3}{Department of Astronomy, University of Michigan, 500 
Church Street, Ann Arbor, MI 48109-1042, USA
}

\email{kwong@ua.edu}


\date{received 2011 May 31; accepted 2011 June 15; published 2011 July 1}

\begin{abstract}

Gas undergoing Bondi accretion onto a supermassive black hole (SMBH) 
becomes hotter toward smaller radii. We searched for this signature 
with a {\it Chandra} observation of the hot gas in NGC 3115, which optical 
observations show has a very massive SMBH.  Our analysis suggests that 
we are resolving, for the first time, the accretion flow within the 
Bondi radius of an SMBH.  We show that the temperature is rising 
toward the galaxy center as expected in all accretion models in which 
the black hole is gravitationally capturing the ambient gas.  There is 
no hard central point source that could cause such an apparent rise in 
temperature. The data support that the Bondi radius is at about 
4\arcsec--5\arcsec\ (188--235~pc), suggesting an SMBH of $2 \times 10^9 
M_{\odot}$ that is consistent with the upper end of the optical 
results.  The density profile within the Bondi radius has a power-law 
index of $1.03^{+0.23}_{-0.21}$ which is consistent with gas in 
transition from the ambient medium and the accretion flow. The 
accretion rate at the Bondi radius is determined to be ${\dot M}_B = 
2.2 \times 10^{-2} \, M_{\odot}$~yr$^{-1}$.  Thus, the accretion 
luminosity with 10\% radiative efficiency at the Bondi radius 
($10^{44}$~erg~s$^{-1}$) is about six orders of magnitude higher than 
the upper limit of the X-ray luminosity of the nucleus.

\end{abstract}


\section{Introduction}
\label{ion_sec:intro}

Understanding accretion onto black holes remains one of the most active 
areas of research in astrophysics today, 
both for probing black hole properties 
and because of their impact on larger-scale problems in 
galaxy and structure formation.  In many nearby galaxies, the model of a 
central black hole gravitationally capturing ambient gas \citep[Bondi 
accretion;][]{Bon52} is a vital part of our understanding of how black 
holes are accreting.

The key to understanding the dynamics of gas in the systems of interest 
lies in correctly modeling the behavior of the accreting gas once it 
falls within the gravitational influence of the black hole, the Bondi 
radius $R_B = 2GM_{\rm BH}/c_s^2$, where $M_{\rm BH}$ is the mass of the black 
hole and $c_s$ is the sound speed of the gas in the vicinity of $R_B$.  
In the absence of angular momentum, the black hole is predicted to 
gravitationally capture the ambient interstellar medium (ISM) 
surrounding it at the Bondi rate
$\dot M_{B} =4 \pi \lambda R_B^2 \rho c_s$, where 
$\lambda=0.25$ for an adiabatic process and $\rho$ is the density of the 
gas at $R_B$.  For observed central $\rho$ values of many nearby 
elliptical galaxies or Sgr A$^\star$, observed luminosities are orders 
of magnitude smaller than what would be predicted based on the Bondi 
accretion rate and assuming a standard 10\% radiative efficiency 
\citep[e.g.,][]{FR95}.  These observational results imply that the 
radiative efficiency of the infalling gas is exceedingly low and/or that 
much less material is accreted by the black hole than the Bondi formula 
implies.

The very weak radiation from most nearby massive black holes has 
prompted a significant theoretical effort aimed at explaining the very 
low radiative efficiencies and/or accretion rates.  The central idea is 
that accretion proceeds via a hot radiatively inefficient accretion flow (RIAF)
rather than a canonical thin accretion disk.  
Early RIAF models include the advection-dominated accretion flow 
\citep[ADAF;][]{Ich77, RBB+82, NY94}, the convection-dominated accretion 
flow \citep[CDAF;][]{NIA00, QG00}, and the advection-dominated 
inflow-outflow solution \citep[ADIOS;][]{BB99}.

Determining which (if any) of these scenarios describes low-$L_X$ black 
hole systems is of fundamental importance to our understanding of 
accretion physics and black hole demography. Observational work has 
focused on using spatially unresolved spectral information to constrain 
theoretical models.  While such studies have been successful in ruling 
out classical ADAF models in some instances \citep[e.g., Sgr 
A$^\star$;][]{DJA+01, Bag+03}, the main limitation has been the 
inability of even {\it Chandra} to resolve the accretion flow inside 
$R_B$ and directly determine the density profiles of the accretion flow, 
as it is the shape of the density profile that most strongly 
distinguishes the theoretical models.  While ADAFs, CDAFs, and ADIOS 
models all predict that the compression of gas within the accretion flow 
within the Bondi radius leads to a $T(R) \propto R^{-1}$ relation for 
the gas, the models predicted significantly different radial density 
profiles (Section~\ref{sec:discussion}).  

In this Letter, we present
for the first time spatially resolved temperature and density profiles
within the Bondi radius of a supermassive black hole (SMBH).
Dynamical modeling of the stellar kinematics of the bulge of NGC 3115 
has confirmed that it harbors a 1--2 $\times 10^9 M_{\odot}$ black 
hole \citep{Kor+96, EDB99}.  At a distance of 9.7 Mpc \citep{TDB+01}, 
this makes NGC 3115 the nearest $>10^9 M_{\odot}$ black hole. Despite 
its large mass, the black hole is very radio quiet, with a core 5 GHz 
luminosity of less than $4 \times 10^{25}$ erg s$^{-1}$ Hz$^{-1}$ 
\citep{FVF98}, several orders of magnitude lower than other black holes 
of similar mass.  The black hole is also undetected in X-rays, with a 
{\it Chandra} upper limit of less than $10^{38}$ erg s$^{-1}$ 
(Diehl \& Statler 2008; K.-W. Wong et al.\ 2011, in preparation).  
The estimated Bondi accretion rate based on the previous observed 
density of $\approx 0.1$ cm$^{-3}$ \citep{DJF+06} in the central part of 
the galaxy is $\sim 0.1 \, M_\odot \, {\rm yr^{-1}}$, implying an expected 
luminosity $\sim 10^{45}$ erg s$^{-1}$ if accretion occurred with $10 
\%$ efficiency.  
The 
lack of radio power argues that it is unlikely that jets are interacting 
strongly with the gas inside $R_B$.

\section{X-ray Observations and Spectral Analysis}
\label{sec:obs}

\begin{figure}
\includegraphics[width=3.truein]{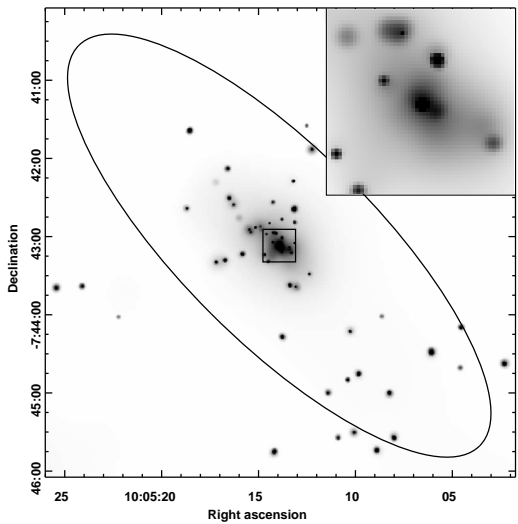}
\caption{
Smoothed 0.3--6.0 keV {\it Chandra} image of NGC3115 with
$D_{25}$ \citep{ddC+91} optical contour. The inset is the 2.0--6.0 keV image of the
central 25$^{\prime\prime}\times 25^{\prime\prime}$ region. There is not
a point source at the center, but a plateau in the diffuse X-ray surface
brightness.
}
\label{fig:image}
\end{figure}

NGC 3115 (Figure~\ref{fig:image}) was observed with {\it Chandra} on 
2001 June 14, 2010 January 27, and 2010 
January 29 for 37, 41, and 77~ks, respectively.  All the 
data were reprocessed using the {\it Chandra} Interactive Analysis of 
Observations (CIAO) software version 4.3 and the {\it Chandra} Calibration 
Database (CALDB) version 4.4.1.  
We used a
local background that is extracted from the 70\arcsec--90\arcsec\ 
annulus centered at the X-ray peak.  The background region is far enough 
from the center so that the source removed surface brightness of the 
X-ray emission is basically flat beyond 70\arcsec.  The background only 
contributes less than about 1\% of the emission (0.5--6.0~keV) within 
the Bondi radius ($ \lesssim 5\arcsec$) and less than about 7\% for the 
ambient region beyond the Bondi flow (4\arcsec --10\arcsec).
Errors are given at $1\sigma$ level unless otherwise specified.

We extracted spectra in annuli centered at the central peak of the 
extended X-ray emission.  
This peak was detected as a source 
by CIAO {\it wavdetect}.  
However, further analysis shows that a modeled 
circular Gaussian width of this peak emission is twice extended as that 
of a nearby bright point source and that the central source is static 
(varied less than 2$\sigma$) within and between the observations. 
Thus, it is dominated by a concentration of hot gas and unresolved 
binaries rather than a single compact accreting object.  We included 
this peak emission but excluded all the other sources detected
by {\it wavdetect} during the analysis.

The unresolved X-ray emission is mainly contributed by unresolved 
low-mass X-ray binaries (LMXBs), cataclysmic variables and coronally 
active binaries (CV/ABs), and the diffuse hot gas component that we are 
interested in. The very soft emission from the gas and the very hard 
emission from the unresolved LMXBs are easy to separate spectrally.  No 
appreciable emission is expected from the gas above 2 keV, so we can use 
the 2--6 keV emission as a lever arm for estimating the amount of 
unresolved LMXB flux in the 0.5--2.0 keV band and subtract it 
accordingly if we assume an appropriate spectral model for the LMXBs.  
\citet{IAB03} have shown that the summed spectra of resolved low-$L_X$ 
($<10^{37}$ erg s$^{-1}$) LMXBs in the bulge of M31 are very similar to 
more luminous LMXBs.
Hence, it is valid to assume that the unresolved LMXB
emission can be spectrally modeled as the brighter resolved sources.
We found that the fainter, softer source like those that make up the 
Galactic Ridge emission (CV/AB) contributes similarly to the soft flux 
of the gas based on $L_X$--$L_K$ scaling relations within $4\arcsec$ for 
this component, where $L_K$ is the Two Micron All Sky Survey (2MASS) 
$K$-band luminosity.  Hence,
including this component is essential in the analysis.

For the LMXB component, we take $\Gamma_{\rm LMXB} = 
1.61^{+0.02}_{-0.02}$ measured from the combined spectrum of all the 
resolved point sources within $D_{25}$ of NGC 3115.  
For the CV/AB component, we fitted the unresolved X-ray emission of M32, 
which is believed to be truly gas free \citep{RCS+07, BKF11}, 
using an absorbed thermal + power-law model.  
The best-fit temperature is $T_{\rm 
CV/AB}=0.59^{+0.05}_{-0.13}$ keV and the power-law index is $\Gamma_{\rm 
CV/AB} = 1.93^{+0.09}_{-0.09}$, 
which are consistent with the values measured by \citet{RCS+08}.  The 
CV/AB normalizations of each annulus are determined by the $L_X$--$L_K$ 
scaling relation derived from M32.

We used the X-ray Spectral Fitting Package\footnote{http://heasarc.nasa.gov/xanadu/xspec/} (XSPEC) for 
spectral analysis.
The 2001 and 2010 spectra of each annulus in the 
0.5--6.0~keV energy range were fitted jointly to the three-component 
absorbed model (PHABS in XSPEC)---a thermal model (APEC in XSPEC) for 
the gas, a power-law (POWERLAW in XSPEC) for the unresolved LMXBs, and a 
combination of thermal +
power law (APEC + POWERLAW) model for the CV/ABs. The absorption is 
fixed at the Galactic value of $N_{\rm H} = 4.32 \times 10^{20}$~cm$^{-2}$ 
\citep{DL90}.  We fit the temperature of the thermal gas component.  
The normalizations of the thermal gas and the LMXB components are 
allowed to vary and untied.  
Since the abundance cannot be constrained from the spectra, we fixed it 
to solar value.  Using 0.2 solar only increases the best-fit temperature 
by $\lesssim 10\%$ and increases the gas normalizations by a factor of 
four.  The derived density only increases by a factor of two and all 
our conclusions remain the same.
All the other parameters are fixed as described above.
The spectra were fitted using the $c$-statistic.
Varying $\Gamma_{\rm LMXB}$ or $\Gamma_{\rm CV/AB}$ by their $1\sigma$ 
uncertainties or varying the CV/AB normalizations by $\pm 20\%$ (that is 
larger than their $1\sigma$ uncertainties) only changes the best-fit gas 
temperature or the gas normalizations by $<$9\%.

\section{Temperature Profile}
\label{sec:tprofile}

\begin{figure}
\includegraphics[width=2.7truein, angle=270]{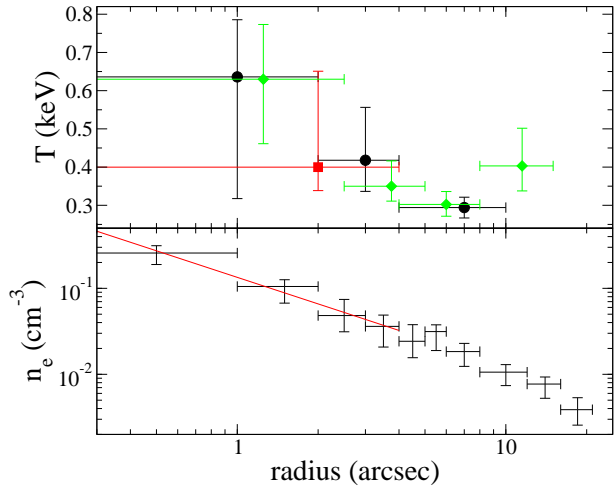}
\caption{
Upper panel: projected temperature profile of the hot gas component.  Different colors are with different binnings. 
Lower panel: deprojected density profile.
The best-fit power-law model within $4\arcsec$ is shown in solid line.
}
\label{fig:t_n_profile}
\end{figure}

The projected temperature profile is shown in the upper panel of 
Figure~\ref{fig:t_n_profile}.  Although the error bars are quite large, 
it can be seen that the trend is rising toward the center as expected in 
a Bondi flow.  
We have also constructed a physical model with a 
temperature profile expected in a Bondi flow (i.e., $T\propto R^{-1}$) 
and a density profile taken from our best-fit model ($\rho(R) \propto 
R^{-1}$; see Section~\ref{sec:SB} below).  The simulated projected 
temperature profile is consistent with our measured profile.  In 
particular, the simulated projected temperature within the 2\arcsec\ 
region is biased low with respect to the deprojected (physical) 
temperature and is consistent with our measured low value.
Thus, the bias in the central temperature may be explained by the 
projection of the cooler gas from the outer region.

\begin{figure}
\includegraphics[width=2.7truein, angle=270]{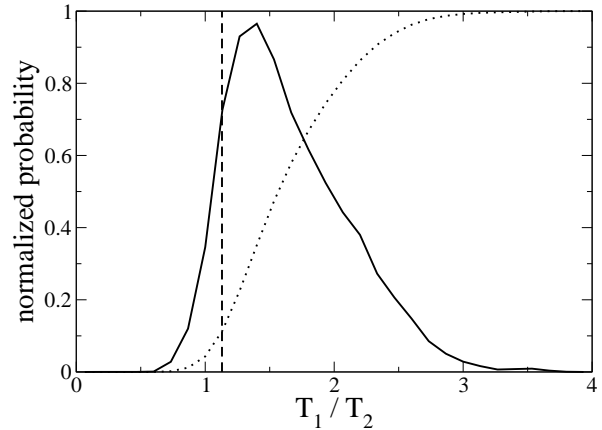}
\caption{
Normalized probability (solid line) and the cumulative probability
(dotted line) of the simulated temperature ratios between the
$2\arcsec$--$4\arcsec$ ($T_1$) and the 
$4\arcsec$--$10\arcsec$ ($T_2$) annuli.  
The vertical line cuts the cumulative probability at 10\%.  }
\label{fig:MCMC}
\end{figure}

For a measured ISM temperature of 0.3 keV in the outer most bin, this 
corresponds to $R_B$=112--224 pc (2\farcs4--$4\farcs8$) depending 
on the adopted black hole mass estimate (1--$2\times 10^9 M_{\odot}$). 
Careful selection of the boundary radius suggests that the onset in the 
increase in temperature begins at $4\arcsec$--$5\arcsec$, rather than 
at $2\arcsec$.  To confirm this, we have performed a Markov chain Monte 
Carlo (MCMC) simulation to
address the significance of the rise in temperature just within a 
radius of $4\arcsec$.  The distribution (normalized probability) and 
the cumulative probability of the simulated temperature ratios between 
the $2\arcsec$--$4\arcsec$ and the $4\arcsec$--$10\arcsec$ annuli are 
plotted in Figure~\ref{fig:MCMC}.  The MCMC test suggests that the 
temperature ratio is larger than 1.13 at the 90\% confidence level. 
Assuming that the temperature beyond $R_B$ is constant and $T \propto 
R^{-1}$ inside $R_B$\footnote{Note that the temperature profile can be 
flatter in the transition region between the ambient ISM and the 
central accretion flow near $\sim 0.1$--$1\, R_B$ \citep{Qua02}.}, the 
expected temperature ratio at 3\arcsec\ and at $R_B$ (4\arcsec) is 
1.33, which is consistent with the peak ratio of 1.4 in the MCMC 
simulation.

The harder spectrum within 
$4\arcsec$ is not the spurious result of any hard central point source. 
We obtained the same 
overall result even when we excised the inner $1\arcsec$.  The effect of 
unresolved X-ray binaries has been properly accounted for by 
quantifying the unresolved emission in the 2.0--6.0 keV band and 
extrapolating to lower energies.
The effect of a large number of softer accreting white dwarfs, stellar 
coronae, etc., 
has also been taken into account in our analysis.  
Such an increase in the 
temperature is exactly what is expected in all models in which the black 
hole is gravitationally capturing the ambient gas \citep[see, 
e.g.,][]{QN99, QG00}.  In fact, a similar spike in ISM temperature was 
found at the center of NGC 4649 and was interpreted as confirmation of a 
$>$$10^9 M_{\odot}$ black hole \citep{HBB+08}, although in this galaxy 
$R_B$ is unresolved by {\it Chandra}. Finally, the lack of detectable 
radio emission from the core of NGC 3115 argues against heating of the 
gas from jets.

\section{Surface Brightness and Density Profiles}
\label{sec:SB}

\begin{figure}
\includegraphics[width=2.7truein, angle=270]{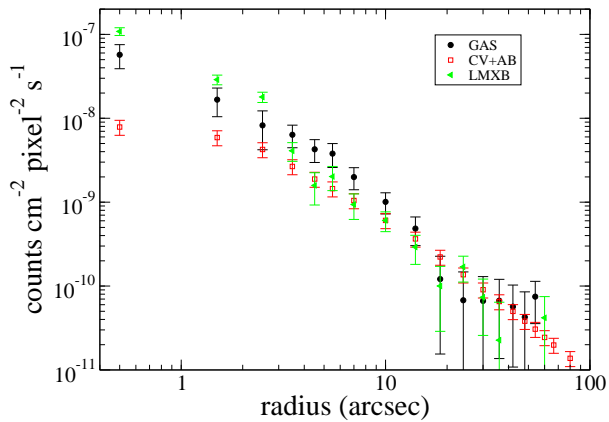}
\caption{
Surface brightness profile for the hot gas 
(black circles), CV/AB (red open squares), and LMXB (green triangles)
components in the 0.5--1.0~keV band.
}
\label{fig:SB}
\end{figure}

We have calculated the surface brightness profiles for the thermal gas, 
CV/AB, and LMXB components.  The CV/AB profile was calculated using the 
same $L_X$--$L_K$ scaling relations mentioned above.  The LMXB profile 
was calculated by assuming that the emission above 2~keV is contributed 
by the LMXBs only, and its contribution to the soft emission was 
calculated by the assumed spectral model above.  The gas profile in the 
soft band was calculated by subtracting the CV/AB and LMXB 
contributions. Figure~\ref{fig:SB} shows the surface brightness profiles 
for the three components in the 0.5--1.0~keV energy band.  
The upper limit is chosen to maximize the 
contribution from the gas component with respect to the other two 
components.  Note that the gas component is clearly detected within 
$\sim 15\arcsec$ and is rising more steeply compared to the stellar 
(CV/AB) component toward the center.

With the surface brightness profile for the hot gas component, the 
electron density profile can be deprojected.  We converted the count 
rate in each annulus into flux using the XSPEC APEC model with a 
temperature of 0.45~keV and solar abundance.  The temperature was taken 
as the mid value of the best-fit temperature of the central 
0\arcsec--2\arcsec\ bin and the ambient temperature.  Since the 
emissivity mainly depends on density, the precise value of temperature 
is not important at our level of accuracy.  
We have tested the density profiles derived using different temperatures 
between 0.3--0.6~keV.  All the derived density profiles are 
consistent within the error bars.
The flux was then deprojected into emissivity, $\epsilon $, at each 
radius using a method described in \citet{WSB+08} and also described in 
\citet{KCC83}.  This technique calculates the emission measures of each 
spherical shell starting from the outer most annulus toward the center, 
and the emission measure of each subsequent shell is calculated by 
subtracting the projected emission from the outer shells.  The electron 
density can then be converted from the emissivity by the relation 
$\epsilon = n_e^2 \Lambda$, where the emissivity function $\Lambda$ 
depends on the temperature and the abundance. The emissivity function is 
taken from the same APEC model.

The deprojected electron density profile is shown in the lower panel of 
Figure~\ref{fig:t_n_profile}.  The errors were estimated by running 
$10^6$ Monte Carlo simulations.  Fitting the density profile within 
$4\arcsec$ to a power law gives $\rho \propto 
R^{-[1.03^{+0.23}_{-0.21}]}$. This is consistent with the power law 
index of $1.14^{+0.29}_{-0.28}$ in the $4\arcsec$--$25\arcsec$ region.

\section{Discussion}
\label{sec:discussion}

Accretion models predict that gas within the Bondi radius flows
toward the SMBH.  The increases in the temperature and density of 
thermal gas toward the center are direct evidence that the gas is being 
gravitationally captured by the SMBH.  Resolving the temperature and 
density profiles within the Bondi radius can provide tight constraints 
to accretion models.

Our 150 ks {\it Chandra} observation of NGC 3115 supports that 
there is a rise in temperature inside $R_B$ compared to the galaxy's ISM 
temperature outside $R_B$, exactly as expected from accretion models.  
We have shown that the onset in the increase in temperature begins at 
$\approx 4\arcsec$--$5\arcsec$ rather than at 2\arcsec.  This supports 
that the mass of the SMBH is at the upper end of the $1$--$2 \times 10^9 
M_{\odot}$ range \citep{Kor+96, EDB99}. 
 
In previous studies where the Bondi accretion flow could not be 
spatially resolved, the overall spectrum has to be modeled to constrain 
accretion model.  With the spatially resolved spectra we have measured, 
we can draw definite conclusion for the first time on the dynamical 
properties within the Bondi radius. In these regions ($R>100R_S$, where 
$R_S=2GM_{\rm BH}/c^2$ is the Schwarzschild radius), the electron 
temperature is fairly well determined since the gas is essentially 
virial and one temperature \citep{QN99}.  Therefore, observations of the 
bremsstrahlung emission in X-ray give direct information on the 
density of the outer regions of the Bondi flow and thereby the accretion 
rate on the outside (${\dot M}(R>100R_S)$).

Because the uncertainties in our measurements are still large, we do 
not attempt to compare the observational results to state-of-art 
theoretical models.  Instead, we focus on the simplest classical 
models. 
In the early analytic 
models \citep[ADAFs;][]{Ich77, 
RBB+82, NY94} it was argued that the radiative efficiency is low because 
the accretion flow is a hot two-temperature plasma in which most of the 
energy is carried into the black hole by the ions.  If the electrons---which 
produce the radiation we observe---receive very little of the 
energy in the system, the radiative efficiency will be much lower than 
the standard value of $\sim$10\%.  In ADAF models, the accretion rate is 
argued to be similar to the Bondi rate and the dynamics of the inflowing 
gas is similar to that of spherical Bondi accretion, even though the 
inflow has angular momentum.
Contrary to ADAF and Bondi predictions, more recent analytic models 
suggest that when the radiative efficiency is low, very little of the 
mass available at large radii actually accretes onto the black hole, 
with most of it being blown away at all radii in an outflow 
\citep[ADIOS;][]{BB99} or 
continuously circulating in convective eddies 
\citep[CDAF;][]{NIA00, QG00}.  
Accretion rates $\ll \dot M_{B}$ are strongly supported by global 
numerical simulations of thick disk accretion \citep[e.g.,][]{IA99, 
SPB99, HB02, INA03}.

Here, we assume the self-similar structure of the accretion 
models extends from the vicinity of the SMBH out to the Bondi radius 
\citep[see, e.g.,][]{AIQ+02}.
In the 
classic Bondi/ADAF models, $\rho(R) \propto R^{-3/2}$, while in CDAFs 
the redistribution of gas via convection predicts a much flatter density 
profile, $\rho(R) \propto R^{-1/2}$.  The ADIOS models allow the 
accretion rate to vary with radius and the density profile is modified 
as $\rho(R) \propto R^{-3/2+p}$, where $p \approx 0$--$1$ characterize 
the density and accretion rate suppression.
In our 150 ks observation, we have constrained 
$p=0.47^{+0.21}_{-0.23}$ from the density profile, which is consistent 
with the value of $p\approx0.6$ measured from the spatially unresolved 
spectrum of Sgr A$^\star$ by \citet{Bag+03}, although they cannot put 
constraints from their data. Taken at face value, we have barely ruled out the 
classical Bondi/ADAF model and the classical CDAF model at about 
$2\sigma$, 
provided that the density profile has a single power-law index within 
$R_B$.

However, theoretical models also suggest that the regions we have 
resolved are still in transition from the ambient ISM to the accretion 
flow \citep{Qua02} and the slope in these regions can be different from 
the asymptotic behavior.  The measured density slope within $R_B$ is 
consistent with that beyond.  Thus, the ADAF/CDAF models might not be 
ruled out.  More theoretical work is needed to interpret the data.  
Nevertheless, the data suggest that the transition does not finish 
until at least $\sim 0.2 R_B$ for either the classical Bondi/ADAF or 
the CDAF models.  The density profile around $\sim 0.1$--$1 \, R_B$ 
does provide some direct constraints to theoretical models, e.g., 
thermal conduction can change the power-law index of the density by 
about 0.5 around those regions \citep{JQ07}.

We determined that the Bondi accretion rate of NGC 3115 is $ {\dot 
M}_{B} = 2.2 \times 10^{-2} \, M_{\odot}$~yr$^{-1}$.
Assuming a 10\% radiative efficiency, the accretion luminosity at $R_B$ 
($10^{44}$~erg~s$^{-1}$) is about six orders of magnitude 
higher than the upper limit of the X-ray luminosity of the nucleus.  
Such a discrepancy can be explained if the accretion rate is lower than 
the classical Bondi rate.  For example, the accretion rate for the 
ADIOS or CDAF models is suppressed and scales with radius as ${\dot 
M}_{\rm acc} \sim \alpha {\dot M}_{B} (R/R_B)^p$, where $\alpha$ is the 
dimensionless viscosity parameter in the standard thin disk model 
\citep{SS73, Bag+03}.  Using a typical value of $\alpha \sim 0.1$ to 
calculate ${\dot M}_{\rm acc}$ at $R_S$, $p$ should be close to 1 to 
explain the discrepancy. It is also quite possible that the gas has 
non-negligible angular momentum that should be more important at 
smaller radii.  This can reduce the accretion rate near the event 
horizon 
(e.g., Proga \& Begelman 2003).  
Alternatively, the radiation efficiency can be lower 
than the canonical value of 0.1 \citep[e.g.,][]{Ho08}.

Currently, the uncertainties are limited by the statistics of our data.  
A deep {\it Chandra} observation will improve the statistics to provide 
useful constraints to accretion models.

\acknowledgments
We thank Eliot Quataert and the referee for helpful discussions. 
This work was supported by NASA LTSA grant NNG05GE48G and {\it Chandra} 
grant GO0-11101A.

\end{document}